# Synchronous control of dual-channel all-optical multi-state switching

**Jiteng Sheng,[1] Junfeng Wang,[1,2] and Min Xiao[1,3,*]**

[1]*Department of Physics, University of Arkansas, Fayetteville, AR 72701, USA*
[2]*School of Physics Optoelectronic Engineering, Nanjing University of Information Science and Technology Nanjing, 210044, China*
[3]*National Laboratory of Solid State Microstructures and School of Physics, Nanjing University, Nanjing 210093, China*
*\*Corresponding author: mxiao@uark.edu*

We have experimentally observed optical multistabilities (OMs) simultaneously on both the signal and generated Stokes fields in an optical ring cavity with a coherently-prepared multilevel atomic medium. The two observed OMs, which are governed by different physical processes, are coupled via the multilevel atomic medium and exhibit similar threshold behaviors. By modulating the cavity input (signal) field with positive or negative pulses, dual-channel all-optical multi-state switching has been realized and synchronously controlled, which can be useful for increasing communication and computation capacities.

All-optical switch, a device utilizing one optical field to control another, i.e., controlling light with light, is an essential component for the next generation of all-optical communication and computing, and has potential applications in quantum information networks. Many approaches have been proposed and demonstrated recently based on various physical mechanisms, such as optical bistability (OB) [1,2], electromagnetically induced transparency (EIT) [3,4], electromagnetically induced absorption (EIA) grating [5], wave mixing [6,7], two-photon absorption [8-14], and enhanced Kerr-nonlinearity [15]. However, most of those approaches are aimed at a binary switch, which has simple on/off states. Quite recently, all-optical multi-state switching has been realized [16] upon a stable control-field-induced optical multistability (OM) [17] in a multilevel atomic medium, where three or more stable output states can appear for a given input state. The advantage of such multi-state switching compared to the ordinary binary switches lies in the increased channel capacity, which ensures more data getting through in the same bandwidth.

In this work, by introducing another optical field, we have observed two outputs with OMs simultaneously on both the signal and the generated Stokes fields in two different output directions of an optical ring cavity containing a hot rubidium vapor cell. We find that the two OMs, which are governed by different physical processes, are strongly coupled via nonlinear processes in the multilevel atomic medium and have similar threshold behaviors, and therefore, the desired states can be reached at the same time in two output channels.

By adding a pulse sequence to the input signal intensity, dual-channel all-optical multi-state switching can be achieve and synchronously controlled, which can be more advantageous in improving the communication capacity [18].

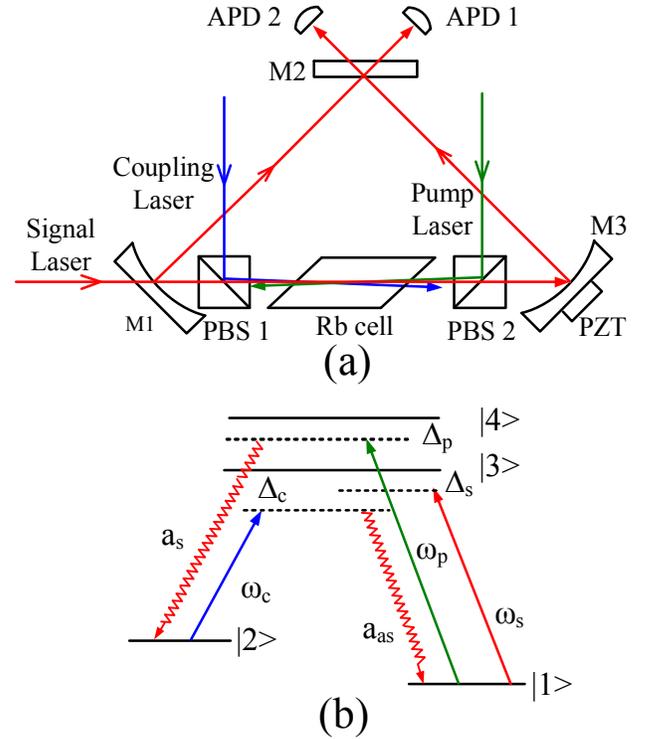

Fig. 1. (Color online) (a) Experimental setup. PBS 1 & PBS 2: polarization beam splitters; M1-M3: cavity mirrors; APD 1 & APD 2: avalanche photodiode detectors; PZT: piezoelectric transducer. (b) The relevant four-level atomic system.

The experimental setup is shown in Fig. 1(a). The signal, coupling, and pump fields are from single-mode diode lasers (Toptica DL 100) with current and temperature stabilized, which operate at the wavelengths of 795 nm, 795 nm, and 780 nm, respectively. A three-mirror optical ring cavity is composed of an input mirror M1 and an output mirror M2 with 3% and 1.4% transmissivities, respectively; and a third mirror M3 with reflectivity larger than 99.5% mounted on a PZT for cavity frequency scanning and locking. The ring cavity length L=37 cm. The naturally abundant rubidium vapor cell is 5 cm long with Brewster windows, and is wrapped in μ-metal sheets for magnetic field shielding and in heating tape for temperature controlling. The signal field is injected into the cavity via the input mirror M1 and circulates in the cavity as the cavity field. The coupling field is injected through PBS 1 and copropagates with the signal field, while the pump field is injected through PBS 2 and counter-propagates with the signal field. Both the coupling and pump fields are not on resonance with the cavity. The two outputs of the ring cavity (through M2) are detected by two avalanche photodiode detectors (APD 1 and APD 2). The radii of the signal, coupling, and pump fields are estimated to be 100 μm, 400 μm, and 400 μm at the center of the atomic cell, respectively. The empty cavity finesse is about 100. The cavity finesse degrades down to about 40 when the atomic cell and PBS are included as the intracavity elements. A fourth laser with the wavelength of ~ 850 nm is used to lock the cavity (not shown in Fig. 1). The relevant energy-level diagram of $^{87}$Rb atoms are shown in Fig. 1(b). The signal, coupling, and pump fields drive the transitions $|1\rangle$-$|3\rangle$ ($5S_{1/2}$, F=1 $\rightarrow 5P_{1/2}$, F'=2), $|2\rangle$-$|3\rangle$ ($5S_{1/2}$, F=2$\rightarrow 5P_{1/2}$, F'=2), and $|1\rangle$-$|4\rangle$ ($5S_{1/2}$, F=1$\rightarrow 5P_{3/2}$, F'=1) with frequency detunings $\Delta_s = \omega_s - \omega_{13}$, $\Delta_c = \omega_c - \omega_{23}$, and $\Delta_p = \omega_p - \omega_{14}$, respectively.

Before introducing OM, we would like to first explain the experimental procedure to observe OB, because either OM or OB can be observed in the same experimental setup by carefully choosing different experimental parameters, for example, the laser frequency detunings, the laser intensities, and the cavity frequency detuning. Figures 2(a) and (b) show the experimentally observed OB curves from two different outputs of the ring cavity propagating in the opposite directions in the ring cavity as recorded by APD 1 and APD 2, respectively. In order to observe two OB curves simultaneously, we first block the pump field, fix the signal and coupling frequency detunings at proper values, and scan the signal intensity triangularly by an electro-optical modulator (EOM) with a relatively slow ramp frequency of ~ 100 Hz, which is similar to the process described in Ref. [16]. A typical OB curve can be observed from APD 2, because OB exists in a broad range of atomic parameter space and is much easier to be observed than OM. Then, the pump field is added, and by tuning the pump frequency detuning and cavity detuning appropriately, two similar OB curves are observed from APD 1 and APD 2 at the same time (Fig. 2).

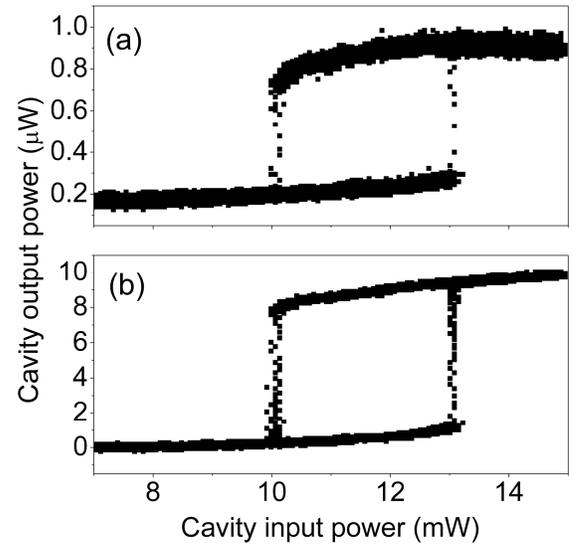

Fig. 2. Experimentally observed OB curves. (a) and (b) are two cavity outputs as a function of input signal intensity recorded by APD 1 and APD 2, respectively. The experimental parameters are $\Delta_\theta$ = 154.7 MHz, $P_c$ = 19.3 mW, $P_p$ = 14.1 mW, $\Delta_s$ = -181.1 MHz, $\Delta_c$ = -208.4 MHz, $\Delta_p$ = -1.2 GHz, and T = 103.2 °C.

Simultaneous OBs at two outputs in a bidirectional optical ring cavity containing a nonlinear medium have been previously predicted by using the mode coupling theory [19]. Qualitatively speaking, in a Doppler-broadened atomic medium and near the atomic resonance, two counter-propagating cavity modes interact with the same group of atoms, and hence the mode coupling can be quite strong which leads to the possibility of having two OBs for the two cavity modes simultaneously. The physical mechanism in the current work is somewhat different from the model used in Ref. [19], because the observed two OBs at the two outputs are from two different physical processes, which can be seen from Fig. 1(b). When both the coupling and pump fields are applied and counter-propagate through the rubidium cell, bright correlated Stokes and anti-Stokes light beams can be generated under appropriate conditions [20,21]. The generated Stokes and anti-Stokes fields travel in the same directions as the pump and coupling fields, respectively. The Stokes field output from M2 is detected by APD 1. Since the fourth laser used for cavity locking (injected through M3) has the same propagation direction as the Stokes field, and also comes out from M2, a grating (Thorlabs GH25-18V) is used to spatially separate the Stokes and cavity locking laser fields. Therefore, APD 1 directly measures the first-order beam of the Stokes field, which is ~ 30 % of the original Stokes field immediately out of the cavity. Due to the overlap between the fourth laser beam and the Stokes field, and the first-order diffraction beam of the Stokes field is directly measured, therefore, the difference between the powers measured by two APDs are as large as 10 times. While APD 2 actually measures both the anti-Stokes and the signal fields, and the signal field is dominant (with the anti-Stokes field less than 30 %). Under the mean-field limit, the steady-state behavior of the output field is governed by $y = \frac{1 - R e^{-i\Delta_\theta}}{T} x - iC\rho$, where x and y are the normalized intracavity and input fields, R (T) is the reflection (transmission) coefficient of the cavity mirrors (with R+T=1), $\Delta_\theta$ is the cavity frequency detuning, $C = \frac{N\omega\mu^2 l}{2\hbar\varepsilon_0 cT}$ is the cooperativity parameter, and $\rho$ is the off-diagonal density-matrix element for the corresponding transition of the field [16,22]. N is the atomic density, $l$ is the length of the atomic sample, $\varepsilon_0$ is the vacuum permittivity, c is the light speed in vacuum, and $\mu$ is the corresponding electric dipole moment. Since the parameters (R, T, $\Delta_\theta$, and C) are all fixed to be constants, then the input-output intensity relation is totally dependent on the expression of the density-matrix element $\rho$ ($\rho$ is also a function of x). When the signal, coupling, and pump fields are all applied on to the same atomic medium, the density-matrix equations can be really complicated and solutions are difficult to obtain even numerically. However, we found that all the density-matrix elements are closely related to the atomic spin coherence term, $\rho_{12}$ [23,24]. Therefore, we can qualitatively understand that all the cavity outputs are determined by the same atomic spin coherence, which is influenced by all the applied laser fields. Thus it ensures that the cavity outputs from two different processes (the Stokes field comes from a nonlinear wave-mixing process and the signal field experiences a near-resonance nonlinear process modified by EIT) have the same OB thresholds, as shown in Figs. 2(a) and (b).

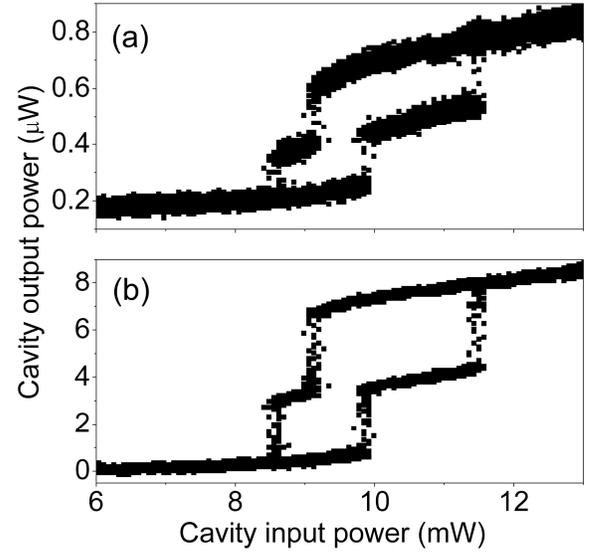

Fig. 3. Experimentally observed OM curves with $\Delta_\theta$ = 181.7 MHz. (a) and (b) are two cavity outputs recorded by APD 1 and APD 2, respectively. Other parameters are the same as in Fig. 2.

OM is very sensitive to the atomic parameters [16], therefore, it requires even more efforts in order to observe two OM curves simultaneously from two outputs of the ring cavity by carefully tuning and searching for the appropriate conditions. Figures 3(a) and (b) depict the observed OM curves from APD 1 and APD 2, respectively. It is worth to point out that by slightly tuning the cavity detuning, e.g., from 181.7 to 154.7 MHz, the OM curves can be transferred back to OB curves (as shown in Figs. 2(a) and (b)). Similarly, the thresholds for both OM curves have the same values.

After the steady OB or OM curves are observed, we fix the signal power in the middle region of the OB or

OM curves and add a pulse sequence onto the signal field with an EOM and a programmable function generator (Tektronix TDS 2014B). A stable dual-channel all-optical binary or multi-state switch can then be practically realized, as shown in Fig. 4, which has been extended from the previous single-channel all-optical switching experiment [16]. From Fig. 4, one can clearly see that the switches for these two channels can be synchronously controlled by only modulating the input signal intensity. Moreover, the wavelengths of the optical fields in these two output channels are different, one at 780 nm and the other at 795 nm, which might be very useful for the future quantum information networks where different characteristic frequencies might be required for different channels. A practical dual-channel multi-state switch requires that the two outputs can be controlled independently, and therefore nine output combinations are possible. However, such function is not available under current experimental conditions, which needs further investigation. We hope that interesting applications can be found for such controllable correlated multi-state outputs with different wavelengths.

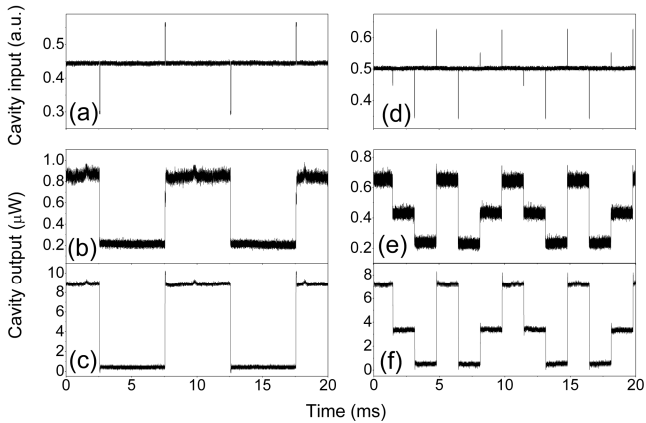

Fig. 4. All-optical binary (left) and multi-state (right) switches. (a) and (d) show the cavity input pulse sequence. (b, c) and (e, f) show the synchronously controlled cavity outputs with two and three stable states, respectively.

In conclusion, we have observed either OBs or OMs simultaneously on both the signal and the generated Stokes fields by scanning the cavity input (signal) intensity, which have outputs from two separated directions of an optical ring cavity with an intracavity hot rubidium vapor cell. Additionally, dual-channel all-optical multi-state switching has been realized and can be synchronously controlled by the input (signal) field with modulated positive or negative pulses. This work opens a new possibility of realizing dual channel multi-state switching for potential applications in information networks and processing with increased capacity.